\documentclass[12pt]{article}
\usepackage{latexsym}
\usepackage{graphics}
\usepackage{epsfig}
\usepackage{epstopdf}
\usepackage{amsmath}
\usepackage{cite}
\usepackage{hyperref}
\usepackage{amssymb}

\begin{document}

\begin{center}
\textbf{\Large Statistical properties of shear deformation of  granular media and analogies with natural seismic processes \footnote{Published in \textit{Pure and Applied Geophysics} (2019) https://doi.org/10.1007/s00024-019-02209-0}} \vspace{0.5 cm}

 Mykulyak S.V.\footnote{e-mail:
\url{mykulyak@ukr.net}}, Polyakovskyi V.V. \footnote{e-mail:
\url{polykovskiyvo@nas.gov.ua}}, Skurativskyi S.I.\footnote{e-mail: \url{
skurserg@gmail.com}} \vspace{0.5 cm}

Subbotin institute of geophysics, Nat. Acad. of Sci. of Ukraine

     Bohdan Khmelnytskyi str. 63-G, Kyiv, Ukraine
\end{center}

\begin{quote} \textbf{Abstract. }{\small 
In this paper, the process of shear deformation of the  medium formed by the cubic  grains  is investigated experimentally. It is demonstrated that the deformation process is intermittent and accompanied by the radiation of acoustic perturbations. These perturbations obey statistical laws that are inherent in  natural seismic processes: the frequency-energy scaling relation (the Gutenberg-Richter law) and  the generalized Omori law for temporal decay of aftershocks. 
The  weak  perturbations influence on the process of shear deformation is studied for the  mono-dispersed medium.
During the granular medium stimulation by weak periodic signals, the existence of  a critical frequency  providing the smallest number of critical events in the granular medium is identified.  We also developed the algorithm  providing  smaller stresses inside the granular massif  by means of external perturbations  during  shear deformation.   Taking into account the statistical similarity between shear and seismic processes, these results open prospects for the ability to affect the natural seismic processes.
}
\end{quote}

\begin{quote} \textbf{Keyword: }{\small 
Granular media; Shear; Self-organized criticality; Seismic processes
}
\end{quote}

\vspace{0.5 cm}
\section{Introduction}
The rock massifs  forming  the lithosphere are extremely diverse, but there is an important characteristics of rocks that is inherent in almost all of them: it is  discreteness  \cite{Sadovskiy1983}.
Discreteness is observed in a wide range of scale levels: for instance, the grains in rocks, the size of which is millimeters or their fraction; the pieces of rock that can be observed in quarries or mountains with centimeters or meters in size; the tectonic blocks having  the dimensions of tens and hundreds of kilometers,  the largest structural elements of Earth's crust -- tectonic plates  extended over thousands, or even tens of thousands of kilometers \cite{Alexeevskaya,Gabrielov_1986,Keilis-Borok,Keilis-Borok_2003,Ben-Zion,Sadovskiy1983}.
Block structurization is most brightly manifested on the boundaries of tectonic plates, where the rock materials were destroyed by earthquakes for a long time \cite{Meade,Billi2003,Billi2004,McCaffrey,Becker,Loveless,Meroz}. 

According to the concept proposed by Sadovsky et al. \cite{Sadovskiy1983}, the geophysical medium is a  thermodynamically open system consisted of hierarchically embedded  blocks. In this system, the long-range spatial and temporal correlations take place causing the emergence of  dynamical  localized structures and other phenomena of self-organization. In particular, the dynamic behavior of the seismically active area is similar to the behavior of a system in the state of self-organized criticality (SOC) \cite{Bak,Sornette,Bak1987}. A number of models were elaborated that reproduce the behavior of seismic areas as being in the SOC state
 \cite{Ito1990,Nakanishi,Barriere1991,Ito1992,Olami,Barriere1994,Baiesi,Kiyashchenko2004,Uritsky2004,Burridge1967,Carlson1989a,Carlson1989b,Christensen1992a,Christensen1992b,Hainzl1999}. In particular, the hierarchical block model taking  the SOC state of seismic region into account  is developed in \cite{Mykulyak2018}.

It is worth noting that similar properties are also endowed with  granular media  under shear deformation \cite{Meroz}. Experimental studies and computer simulation of deformation of granular massifs indicate that fluctuations of intergranular forces and  velocities of granules can many times exceed their mean values \cite{Danylenko2017,Behringer,Howell,Cabalar} that is common to  seismic processes \cite{Sornette} as well. 
During the deformation of granular systems, the long-range correlations \cite{Mykulyak2019} are intensified. Moreover, the displacement of granular media occurs intermittently \cite{Behringer,Howell,Zhao,Indraratna} and is accompanied by the radiation of stochastic acoustic disturbances. 
Investigation of these regularities and granular media responses on external influences are important for understanding the  discrete media behavior taking place at different scales, in particular on the scale associated with the earthquakes generation.

In this paper, we investigate experimentally the shear deformation of granular medium consisted of   cubic  grains. Section \ref{mps:sec2} contains the experimental rig description.  In Section \ref{mps:sec3} the  influence of additional massif load   on the deformation process and  statistical properties of  acoustic emission   released  by  considered granular system are studied in more detail.  In Section \ref{mps:sec4} it is examined   the  shear processes in granular medium when  the periodic and impulse external disturbances are applied. The final section contains the concluding remarks. Notably, from the foregoing provisions it follows that  the behavior of this system is similar to the behavior of the geomedium  in a seismic zone. Therefore, such studies  may help in understanding  the unknown properties of natural seismic processes and possibility of artificial influence on them.

\section{Experimental installation for the study of shear deformation of massif  formed from cubic grains}\label{mps:sec2}

\begin{figure}[bt]
\begin{center}
\includegraphics[width=11 cm, height=6 cm]{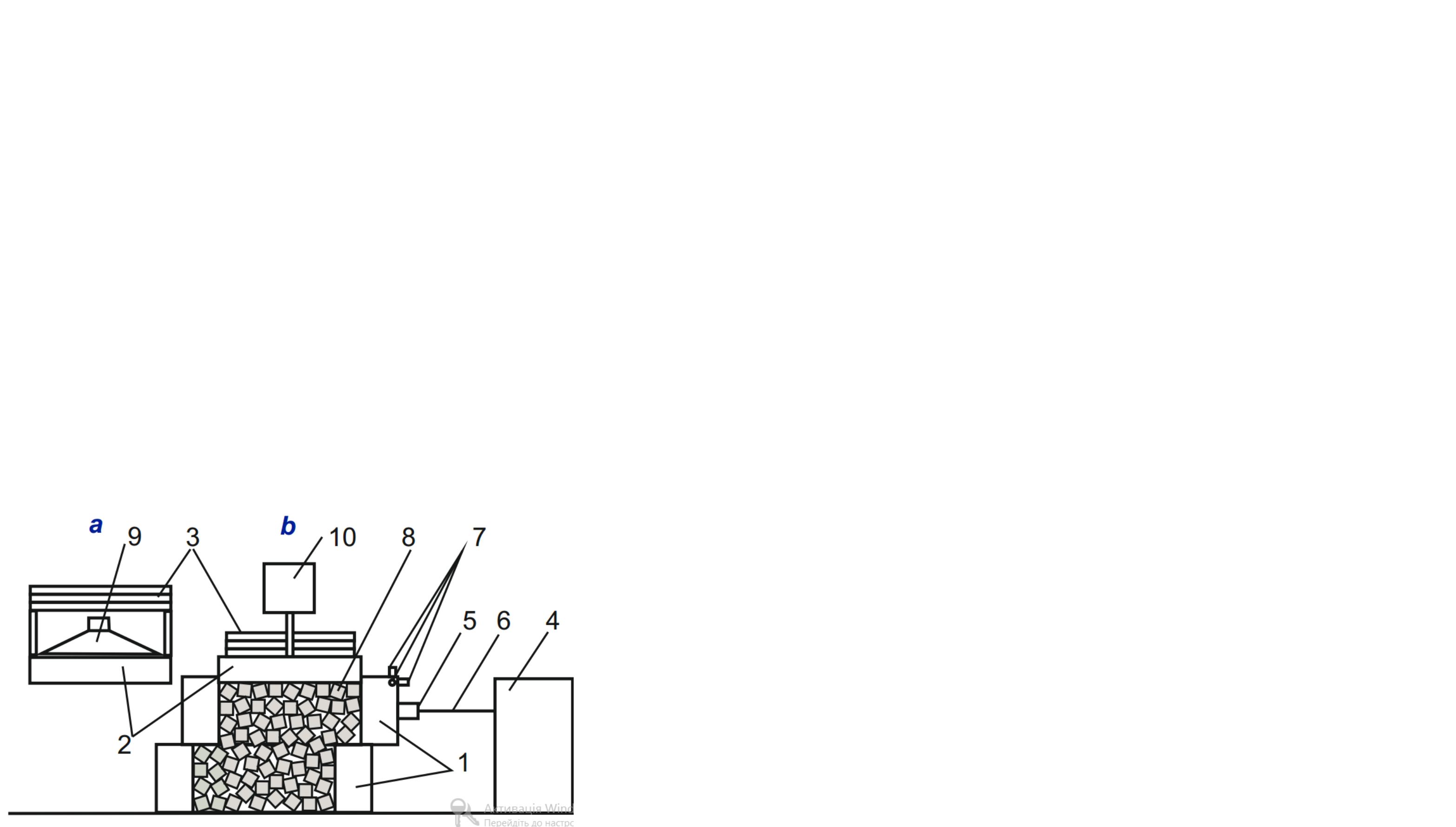}
\end{center}
\caption{The sketch of  experimental setup. Designations : 1 -- box consisting
of two parts, 2 -- piston, 3 -- load, 4 -- traction mechanism, 5 -- force sensor, 6 -- rope, 7 -- accelerometers, 8 -- cubic grains, 9 -- acoustic speaker, 10 -- impulse generator. The module (a) is used in the case of periodic action on the granular medium, whereas (b) is applied for producing the impulse action.
} \label{mps:fig1}
\end{figure}

The study of  shear deformation process of granular media was carried out  for both cases of external perturbations and their absence. The experimental installation is the box, which is made from plexiglass and consists of  lower fixed part and upper slider one (Fig.~\ref{mps:fig1}).  The lower box part has the following internal sizes: the length along shear direction 0.3~m, the  hight  0.07~m, and the width 0.2~m. The front and back wall thickness is 0.1~m.  The slider part is of  the same dimension. 
 The grinding of contact surfaces between lower and upper box parts was performed. 
To provide the directional sliding of the upper part, the guide plates were installed on the lower part.   
In turn, the slider part of the device
is equipped with  the limiters prohibiting  the movement  in the vertical direction. The piston (2) that can move freely in a vertical direction is located on the granular massif.

The movement of the slider block is carried out using a traction mechanism (4) providing the slider block displacement  0.1~m in a minute. This device consists of a gear motor NMRV 090/040, which is connected to the rope shaft. One end of the rope (6) is attached to the shaft, and the other to the force sensor KMB 19 (5), which is tightly tied to the leading edge  of the upper box.

On the slider part  there are three single axis accelerometers  T-500 (7)  measuring the  accelerations in three orthogonal directions.  The force sensor (5) on the  leading edge  measures the granular massif reaction  on the shear deformation.

The study of external perturbations influence  on the granular medium shear is carried out for two types of actions: a) periodic and b) impulse. In the first case, the module (a) consisting of piston (2), load (3) and acoustic low-frequency speaker  75-GDN-4 (9) is used. The speaker generates the periodic weak disturbances with  10 Watts power   in the frequency range of 50-1000 Hz. 

To investigate the impulse perturbations influence  on the shear process, instead of the module (a),  the module (b) is used. The perturbations are generated by an impulse generator (10)  mounted on the upper surface of the piston (2). There is also possibility to vary the static loading (3).

\section{The characteristics  of granular medium reaction   on shear deformation}\label{mps:sec3}

In the  first series of experiments, let us consider how the magnitude of static loading of the granular medium affects  the process of its shear deformation. The experimental set-up is used with the module (a) when the acoustic speaker (9) is turned off. 

Monodisperse granular medium is formed by the massif of 3000 cubic grains made of plexiglass with the following characteristics: Young's modulus is  $E=3\cdot 10^9$~Pa, Poisson's ratio is $\nu=0.3$,  and density  $\rho=1200$~kg/m$^3$. The coefficient of friction between the surfaces of the grains is $k = 0.4$. Initially, the grains fill the volume of the box in such a way that the piston is immersed in the box at the depth of 0.4~mm. Thus, the packing fraction (solid volume fraction) of the granular medium is $\gamma = 0.368$.

According to the estimates \cite{Meroz2017},
 the average size of blocks in the area of San Andreas fault in California is about 90 km and the width of the area, where the main deformation occurs between the tectonic Pacific plate 
the North American plate, is 565 km. Thus, the ratio of the area width   to the block size is about $\kappa \approx 6$. This ratio  belongs to the interval 5-10 of the ratio $\kappa$ observed in experiments with a shear deformation of the granular medium \cite{Howell1999}.
In our case reduced width of the granular massif is about 14, which is close to the shear band width. 
The packing fraction of the monodisperse granular medium is $\gamma = 0.368$. This packing fraction value is less then critical one ($\gamma = 0.5$) when the crystallization appears \cite{Roding2017}. 
Under natural conditions, crystallization is unlikely also, as the discrete medium formed by geoblocks is not monodisperse.

%
%
\begin{figure}[bt]
\begin{center}
\includegraphics[width=5.5 cm, height=5.5 cm]{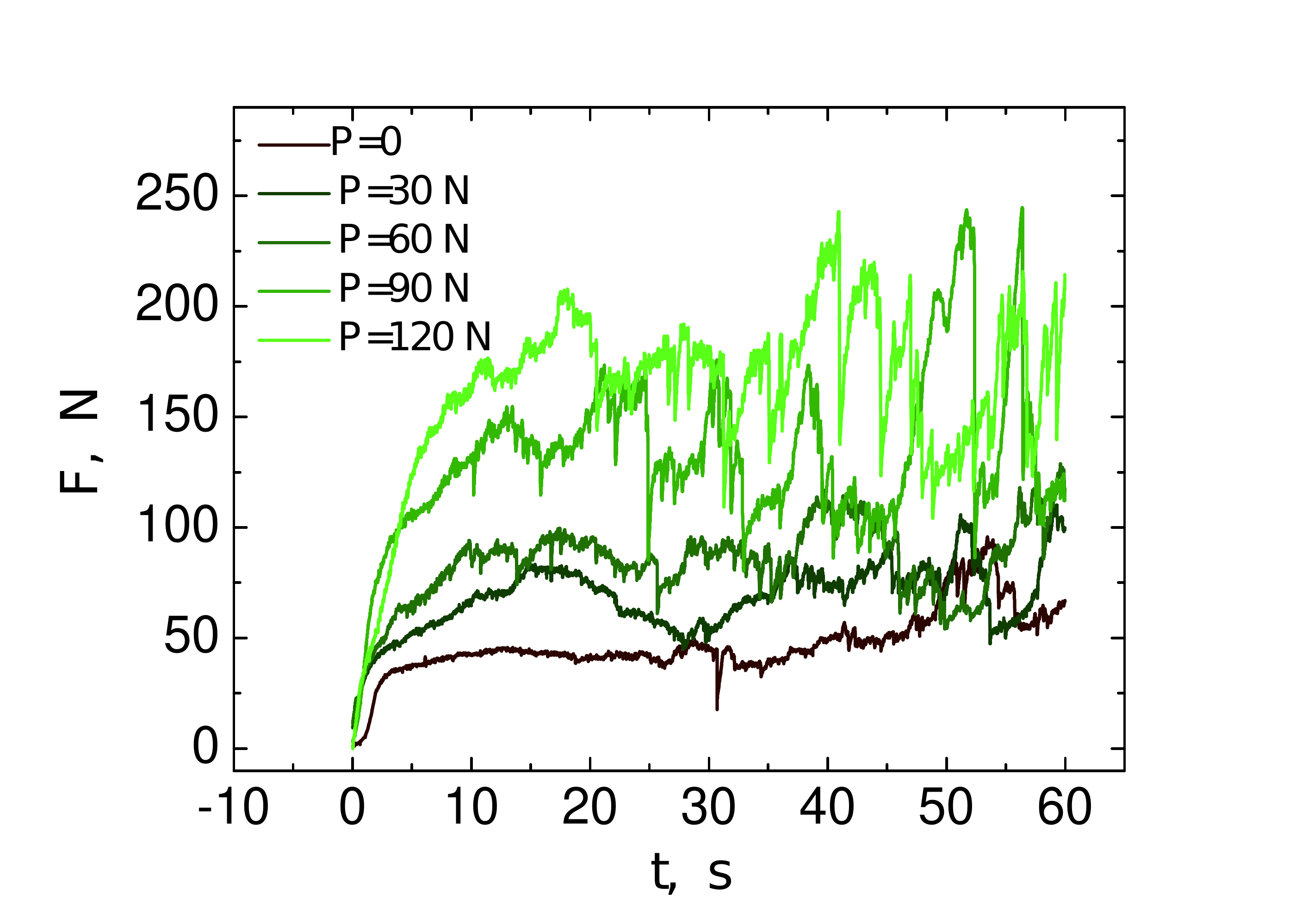}\hspace{0.1 cm}
\includegraphics[width=5.5 cm, height=5.5 cm]{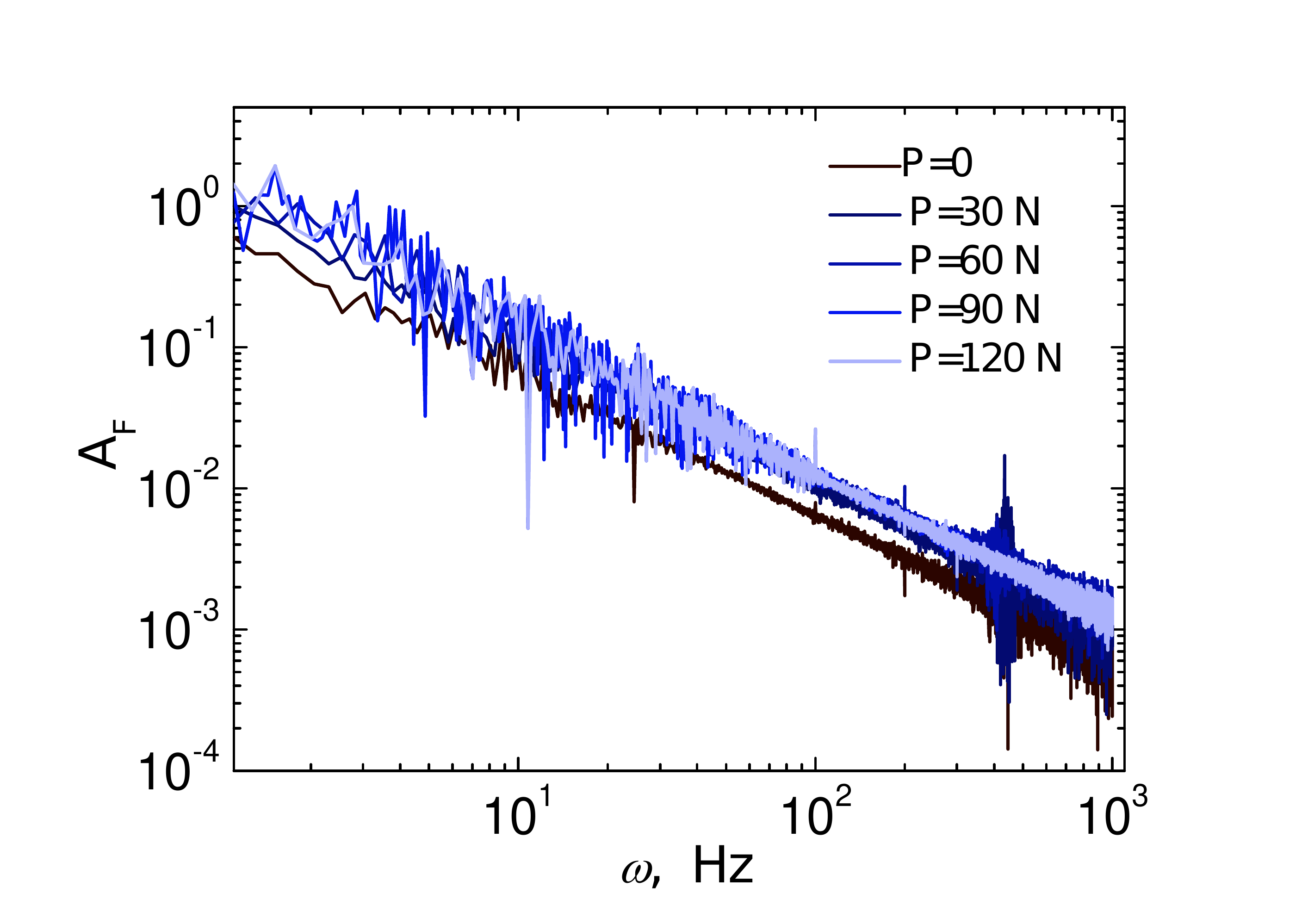}
\centerline{(a) \hspace{5 cm} (b)}
\end{center}
\caption{Temporal dependences (a) of the force $F$ at  loading  $P=0$, $30$, $60$, $90$, $120$~N (curves from bottom to top) and their Fourier spectra (b). 
} \label{mps:fig2}
\end{figure}

\begin{figure}[bt]
\begin{center}
\includegraphics[width=5.5 cm, height=5.5  cm]{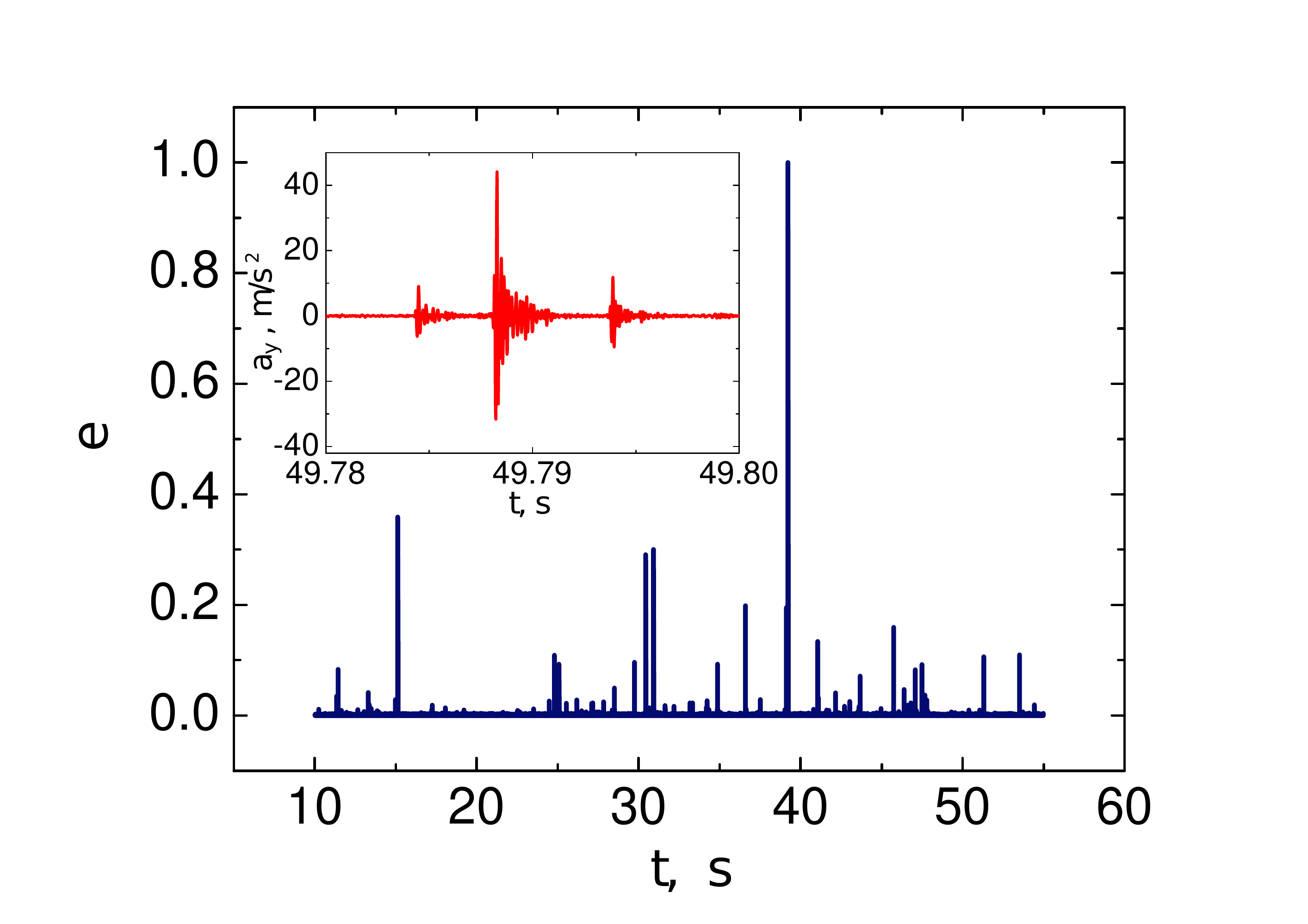}\hspace{0.1 cm}
\includegraphics[width=5.5 cm, height=5.5  cm]{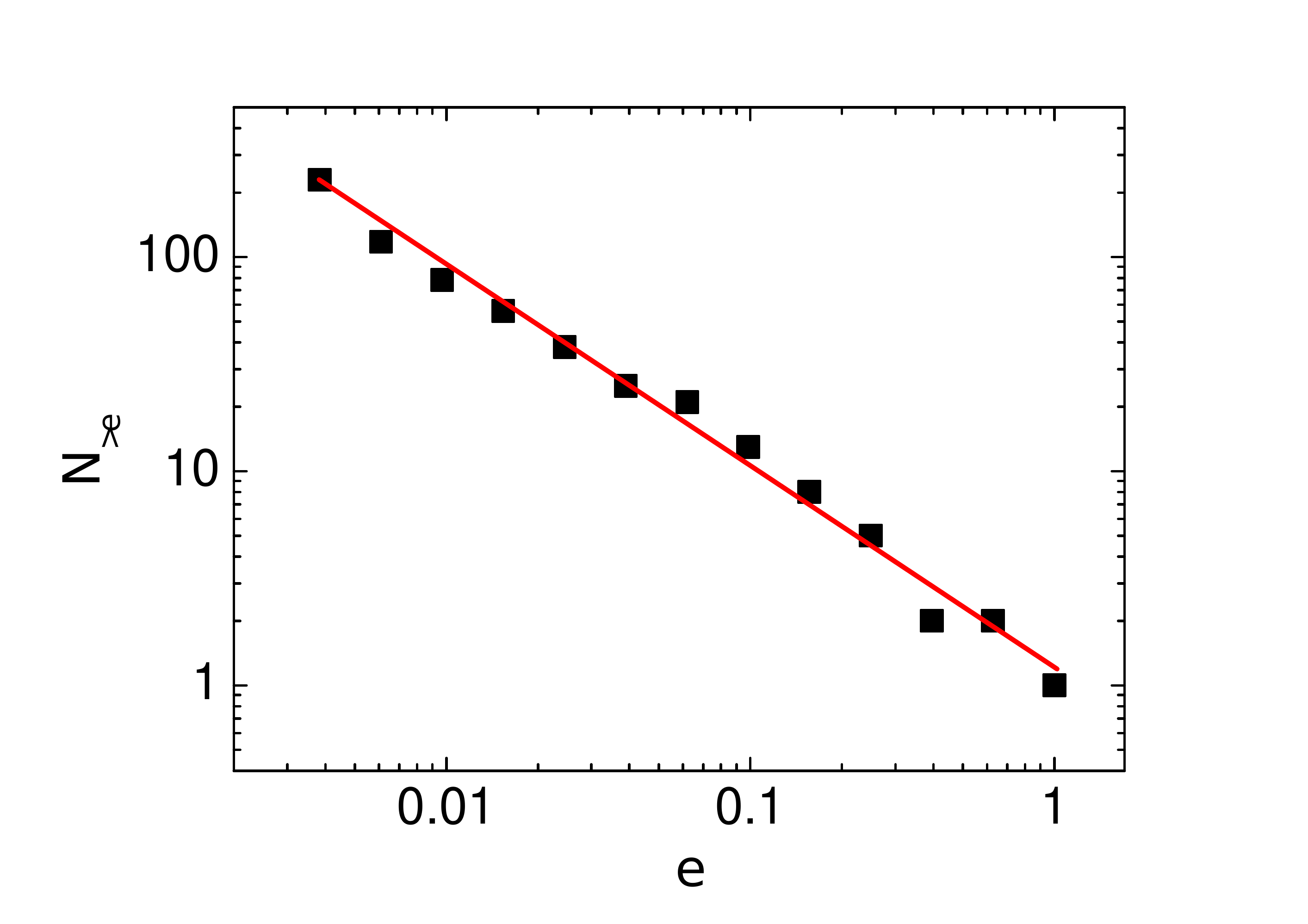}
\centerline{(a) \hspace{6 cm} (b)}
\end{center}
\caption{Sequences (\ref{mps:seq}) of acoustic perturbation energy depending on time (a) and corresponding cumulative complementary  distribution (square dots) of perturbations by energy (b) and its approximation by power relation (solid line). The inset in left panel  is the profile of $a_y$-component for the acceleration. 
} \label{mps:fig3}
\end{figure}
The temporal dependence of the traction force $F(t)$  registered with the force sensor is depicted in Fig.~\ref{mps:fig2}a. Analysis of these dependencies indicates a tendency to increase the force of reaction $F$ with increasing loading on the piston. Performing the discrete Fourier transform of temporal $F(t)$ dependences, the corresponding  spectra of these dependencies are obtained (Fig.~\ref{mps:fig2}b). All spectra  are similar and obey the power relations. The  estimation of the slope of regression line for the data in Fig.\ref{mps:fig2}b gives the degrees of power which are close to each other and are equal to about $-0.998\pm0.003$. This points to the similarity of processes at different loading, as well as to the scale invariance of the shear deformation process.

During   medium deformation the acceleration $a(t)$  in acoustic waves generated by the medium is examined. We measure all components of $a(t)$ with accelerometers (7)
(Fig.~\ref{mps:fig1}). 
The inset in Fig.\ref{mps:fig3}a shows the part of the signal $a_y(t)$ in the interval $t\in[49.78; 49.80]$~s recorded by the accelerometer in the direction of movement of the upper device part. This signal is a sequence of perturbations generated by the granular system. Each component $a_f$, $f=\{x,y,z\}$ is integrated in order to obtain the temporal dependence of the velocity component $v_f$. 

 To distinguish between individual pulses we have used an auto\-regres\-sive\--Akaike information criteria (AR-AIC) picker \cite{Leonard1999,Leonard2000}.

Next, let us introduce the quantities
\begin{equation}\label{mps:seq}
e_i^\star=\frac{1}{t_i^e-t_i^b}\int_{t_i^b}^{t_i^e} \sum_{x,y,z} v_f^2 (t)dt, 
\end{equation}
where $t_i^b$ and $t_i^e$  are the initial and final moments of the $i$th perturbation, i.e. the boundaries of separate pulse (the inset in Fig.~\ref{mps:fig3}(a)). These quantities are proportional to the energies \cite{velocitymean} of  the  perturbations. As a result, we obtain the sequence $e_i=e_i^\star/\max{\left(e_i^\star\right)}$ 
which can be regarded as a sequence of reduced energies of acoustic perturbations. 

In particular, the  sequence $e_i$ for the experiment when  the loading on a box cover is $P=120$~N is shown in Fig.~\ref{mps:fig3}(a). 
Now the question arises: does this sequence  obey the statistical laws 
that take place for  seismic processes?

At first, consider the cumulative complementary distribution of acoustic perturbations for energies. It can be approximated  by the power relation
\begin{equation}\label{mps:eq2}
N(>e)=C e^{-\beta},
\end{equation}
where $\beta=0.93 \pm 0.03$, $C=-4.88 \pm 0.04$.  From Fig.\ref{mps:fig5}(b) it follows that the approximation matches the experimental distribution. Therefore, this distribution is of power nature as   Gutenberg-Richter's law \cite{Gutenberg1944}. Moreover, the index $\beta$ belongs to the interval 0.8-1.05 which is typical for seismic processes \cite{Olami}. 

\begin{figure}[bt]
\begin{center}
\includegraphics[width=5.5  cm, height=5.5  cm]{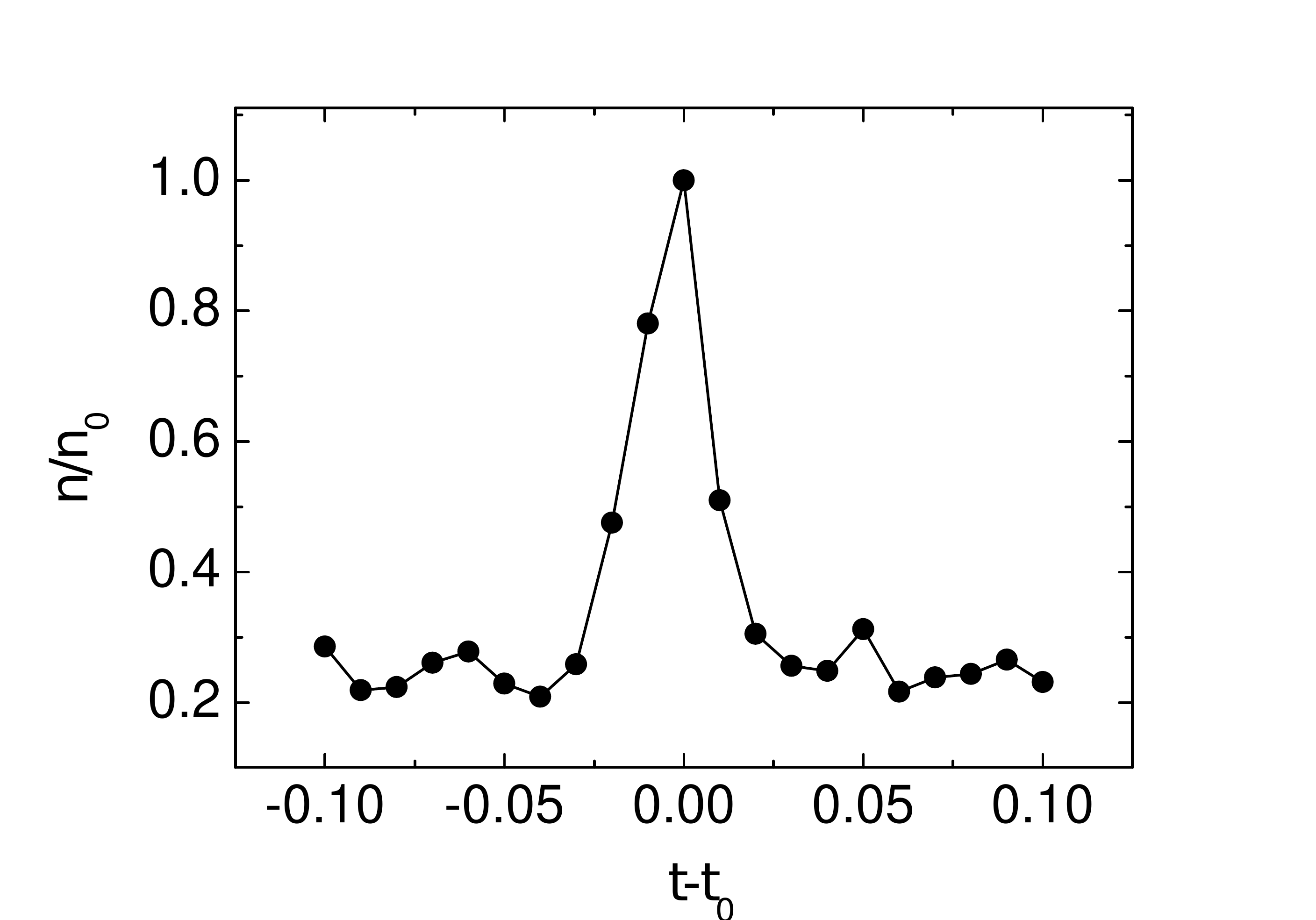}
\includegraphics[width=5.5 cm, height=5.5  cm]{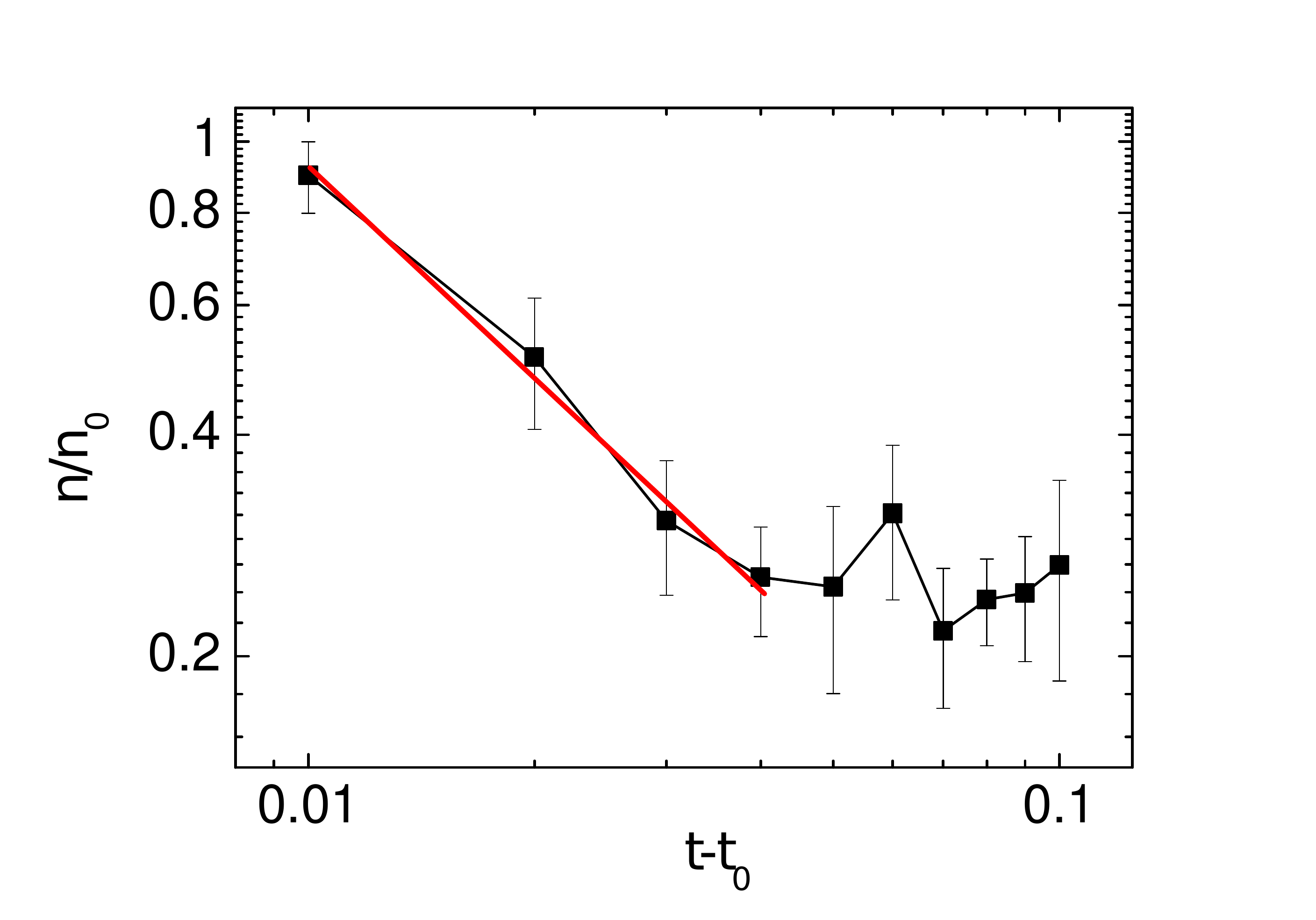}
\centerline{(a) \hspace{6 cm} (b)}
\end{center}
\caption{Dependence of the average number of acoustic disturbances on time to a major earthquake (a); power approximation of aftershocks (b). Here $n_0$ stands for the number of main shocks, $n$ is the total shock number, and the 95\%  confidence intervals are shown.
} \label{mps:fig4}
\end{figure}

The statistics of  foreshocks \cite{Jones1979} and aftershocks \cite{Omori1894} caused by  earthquakes is an important characteristics of seismic zones and their dynamics. Consider analogous phenomena associated with large emission of energy exceeding the threshold $e_{th}=0.02$ (Fig.~\ref{mps:fig4}a). It turns out,  model aftershocks attenuate according to the Omori law  \cite{Omori1894} which is valid for natural aftershocks of large earthquakes \cite{Utsu1969}:
$$
N=\frac{k}{(t+c)^p},
$$
where the coefficients $k=0.13 \pm 0.03$, $c=0.01$, $p=0.91 \pm 0.06$ (Fig.~\ref{mps:fig4}b). The presence of foreshocks and aftershocks testifies to the existence of temporal correlations in the process of generation of perturbations.

\begin{figure}[bt]
\begin{center}
\includegraphics[width=5.5 cm, height=4.5 cm]{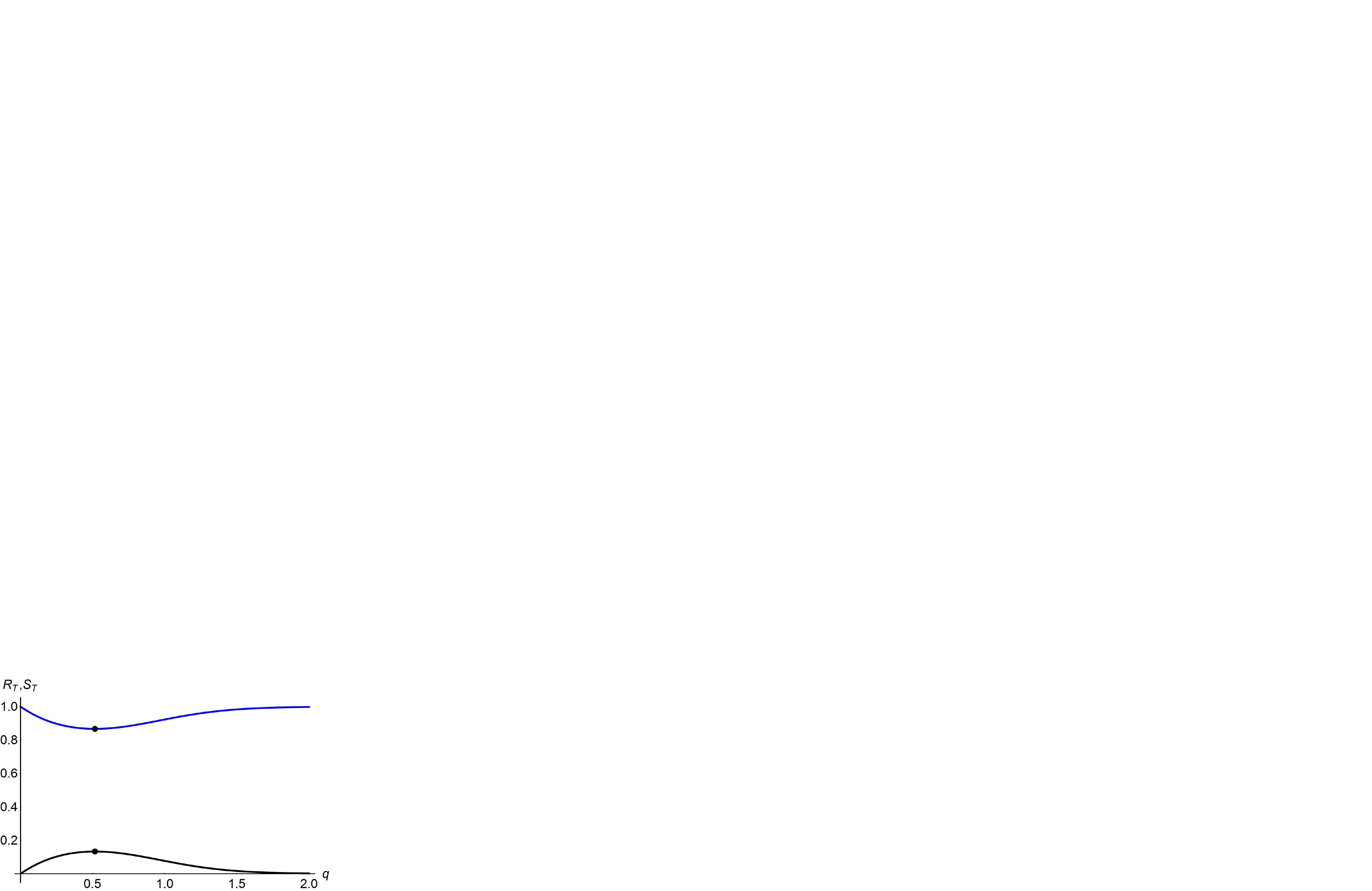}
\end{center}
\caption{The estimation of the entropy index $q$. The functions $R_T$ (upper curve) and $S_T$  (lower curve).
The extremum abscissa  is $q=0.516$.} \label{mps:fig18}
\end{figure}

The nature of the observed experimental data 
can be examined by using the generalized entropy  conception \cite{VallianatosEPL2012,VallianatosAG2012,VallianatosTectphys2013,VallianatosStMAp2013,MichasNPG2013,VallianatosPandAG2014,%
MichasEPS2015,VallianatosETIS2015,AgioutantisIJMST2016,Hloupis2016,%
VallianatosInv2016,Kaklis,VallianatosMA2018,SaltasCSTS2018}. 
This extension of thermodynamics and statistical physics introduced by Tsallis \cite{TsallisIntr} deals with the nonergodic systems and leads to the nonextensive thermodynamics and statistical mechanics. Instead of classical entropy, the quantity 
$S_T$   is defined 
$$
S_T=\frac{1-\sum_{i=1}^Q p_i^q}{q-1}, \qquad q\in R,
$$
where $q$ is the Tsallis (or entropy) index, $\sum_{i=1}^Q p_i=1$. Note that the limit $q \rightarrow 1$ corresponds to the Boltzmann-Gibbs entropy  $S_T\rightarrow -\sum_{i=1}^Q  p_i\ln p_i$ and the deviation of $q$ from 1  points to the  appearance of  long-range correlations.
It has been shown that at $q<1$ the physical system behavior depends on  rare events, whereas at $q>1$ the frequent events have more weight \cite{BoghosianTs}. 

To estimate the index $q$, the approach proposed in \cite{Tsallisq} is used. According to this method, the maximum entropy principal on the base of  $S_T$ 
is applied. The auxiliary function known as redundancy $R_T$ is defined
$$
R_T=1-\frac{S_T}{S_{T\max}},
$$
where $S_{T\max}=\frac{1- Q^{1-q}}{q-1}$ is the maximum of the function $S_T$  which  is reached on the equiprobable microstates $p_i=1/Q$. 

We thus construct the functions $S_T$ and $R_T$  using the sequence $e_i$ presented in 
Fig.~\ref{mps:fig3}a. Each of these functions possesses single extremum depicted in Fig.~\ref{mps:fig18} with the filled circles.  The numerical estimation of extremum coordinates gives $q=0.516<1$. This means that the process described by this distribution possesses long-range correlations and the system dynamics is defined by the mutual influence of a large number of rare events.

\begin{figure}[bt]
\begin{center}
\includegraphics[width=7 cm, height=6 cm]{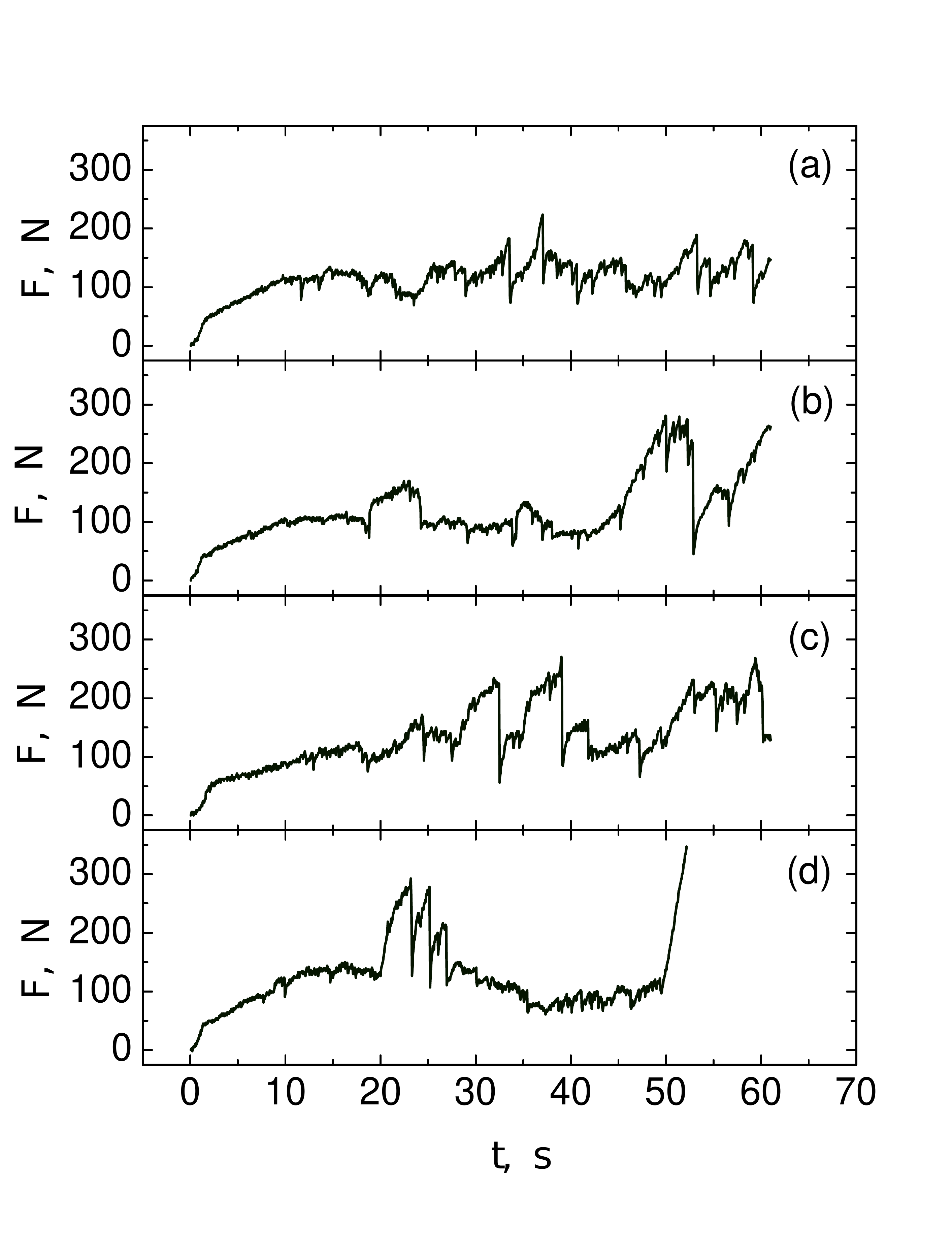}
\end{center}
\caption{
Temporal dependences for the traction force $F$ when the medium is monodisperse: identical cubes of size 10 mm (a) and polydisperse: the mixture of cubes of sizes 10 mm and 25 mm in proportions: b -- 80\% $\times$ 20\% (number of grains: 2400$\times$38), c -- 70\%$\times$30\% (2100$\times$59), d -- 50\%$\times$50\% (1500$\times$96).
} \label{mps:fig5}
\end{figure}

The next series of experiments is concerned with the study of  influence of medium's heterogeneity  on  its deformation. To do this, 
the  mixture of cubic grains of two sizes $l_1=10$~mm and $l_2=25$~mm is used. The shear deformation performs at loading $P=60$~N. Temporal dependences of the traction force (Fig.~\ref{mps:fig5}) show that the process of deformation of both monodisperse and disperse media is not significantly different. 
This  is also confirmed by the comparison of their Tsallis indexes which are practically unchanged for the mixtures of cubes.
However, when number of grains with the edge $l_2=25$~mm  increases,  the  stress increases as well. Note that if the bulk ratio of cubes of different volumes equals 0.5, the significant increase in traction  force takes place leading  to  the rope destruction   (Fig.\ref{mps:fig5}d).

\section{Influence of external perturbations on shear deformation}\label{mps:sec4}

According to the studies presented above, the behavior of the granular medium in the process of shear deformation is enriched by  complicated stochastic reactions to  shear loading. This response of the system obeys the statistical laws that are inherent in the natural seismic process. The question arises whether it is possible to make a change in the behavior of this complex system with the help of small disturbances?

As above, let us consider the shear deformation of the granular medium formed by 3000  cubes of the size $l_1=10$~mm. The displacement is carried out with the device described in Fig.\ref{mps:fig1}. The \emph{periodic} perturbations of the 50-1000~Hz frequency range  are injected into the medium with the help of  acoustic speaker mounted in the module (a). The massif is stressed by the load $P=60$~N.

\begin{figure}[bt]
\begin{center}
\includegraphics[width=5.5 cm, height=5.5  cm]{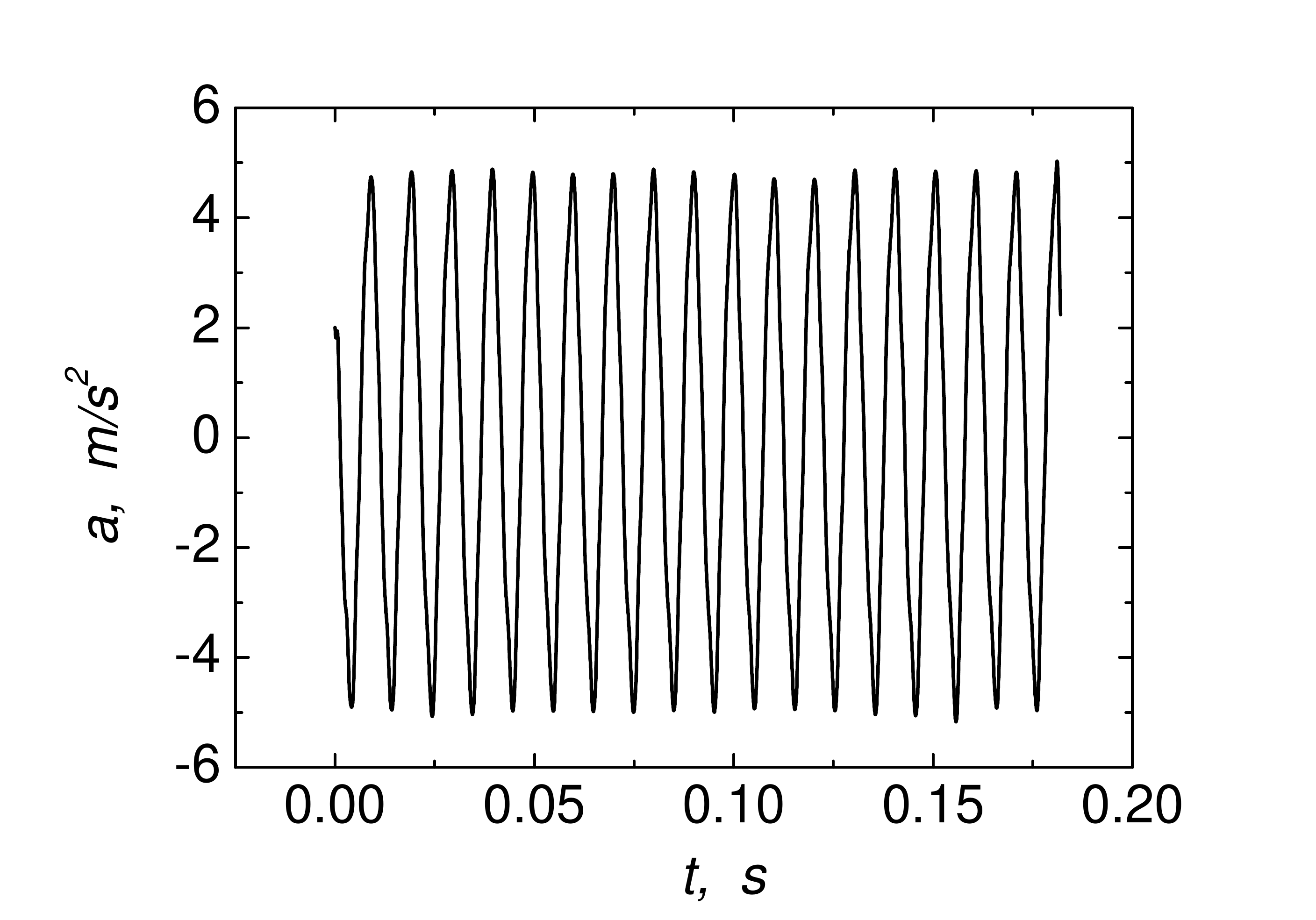}\hspace{0.1 cm}
\includegraphics[width=5.5 cm, height=5.5  cm]{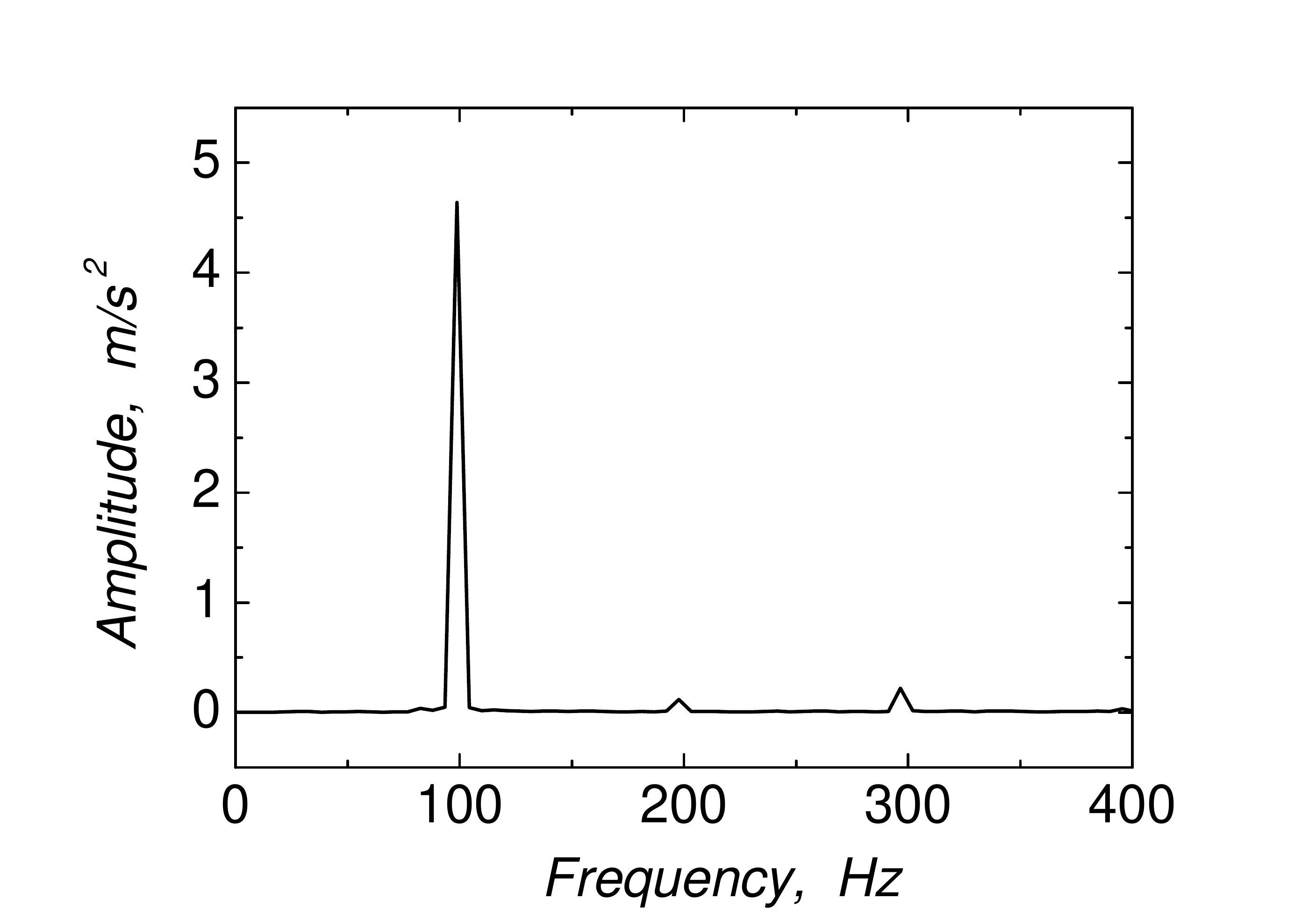}
\centerline{(a) \hspace{6 cm} (b)}
\end{center}
\caption{The signal recorded by the accelerometer at the bottom of the piston (a) and its spectrum (b).} \label{mps:fig6}
\end{figure}
\begin{figure}[bt]
\begin{center}
\includegraphics[width=5.5  cm, height=5.1  cm]{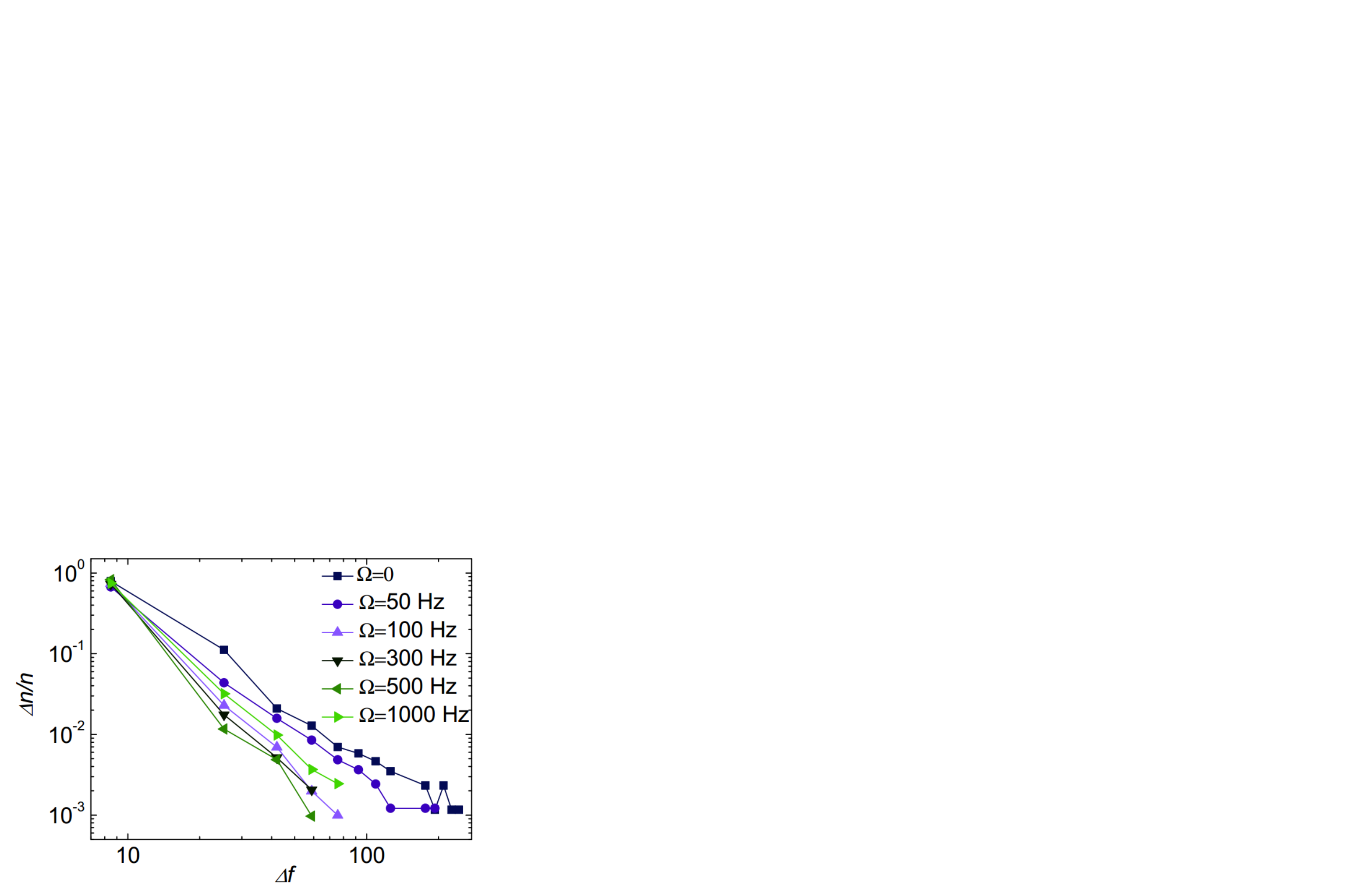}\hspace{0.1 cm}
\includegraphics[width=5.5  cm, height=5.5  cm]{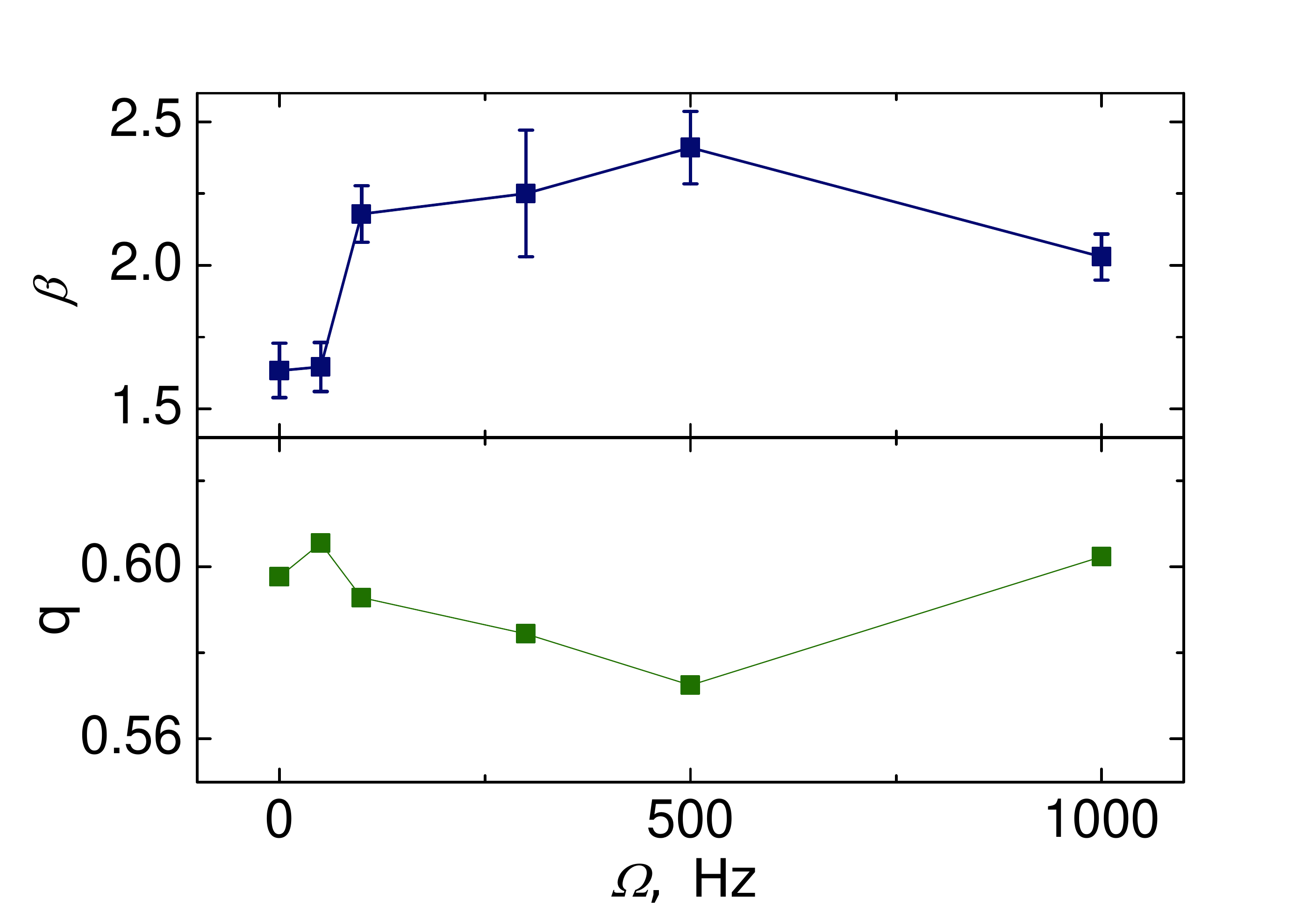}
\centerline{(a) \hspace{6 cm} (b)}
\end{center}
\caption{The distribution of force jumps at different values of perturbation frequencies $\Omega$ (a), the power index $\beta$ (upper panel)  in the  approximation of distributions and Tsallis index $q$ (bottom panel) versus  the frequency  of perturbation $\Omega$ (b).
} \label{mps:fig7}
\end{figure}
To check what kind of  signal is entered the medium, an additional accelerometer  is installed in  the bottom of the piston. Fig.~\ref{mps:fig6}a shows the accelerations, measured by this sensor, and  its Fourier spectrum  at the  100~Hz frequency of   input periodic signal (Fig.~\ref{mps:fig6}b). There is the main maximum in the diagram, whereas the spectral amplitudes of other harmonics are insignificant. Similar spectra have signals received for input  periodic perturbations at frequencies of 50, 300, 500, 1000~Hz. From this it follows that the signal passing through the piston  is almost not distorted. 

Thus, after supporting explanations, let's return to the process of deforming under the action of traction force $F$. To  analyze  the influence of external perturbations, jumps of force are calculated as the difference between adjacent local maxima and  minima. These jumps of force are associated with the reaction of the block media to the shear. The constructed distributions of the number of jumps at their intensity for the five frequencies of periodic perturbations are shown in Fig.~\ref{mps:fig7}a. From this figure it follows that these distributions  are close to power functions and depend on the frequency of perturbations. The dependence of the power index on the frequency is shown in Fig.~\ref{mps:fig7}b. It turns out that at a 500~Hz frequency  the index has  a local maximum. That is, at this frequency, the number of large jumps 
of force is the smallest. In addition, the maximum values of jumps are the smallest among all
 frequencies.
This allows us to conclude that under these conditions the deformation process is the most  smooth and ''soft'', without sharp jumps of force.
\begin{figure}
\begin{center}
\includegraphics[width=7 cm, height=5.5 cm]{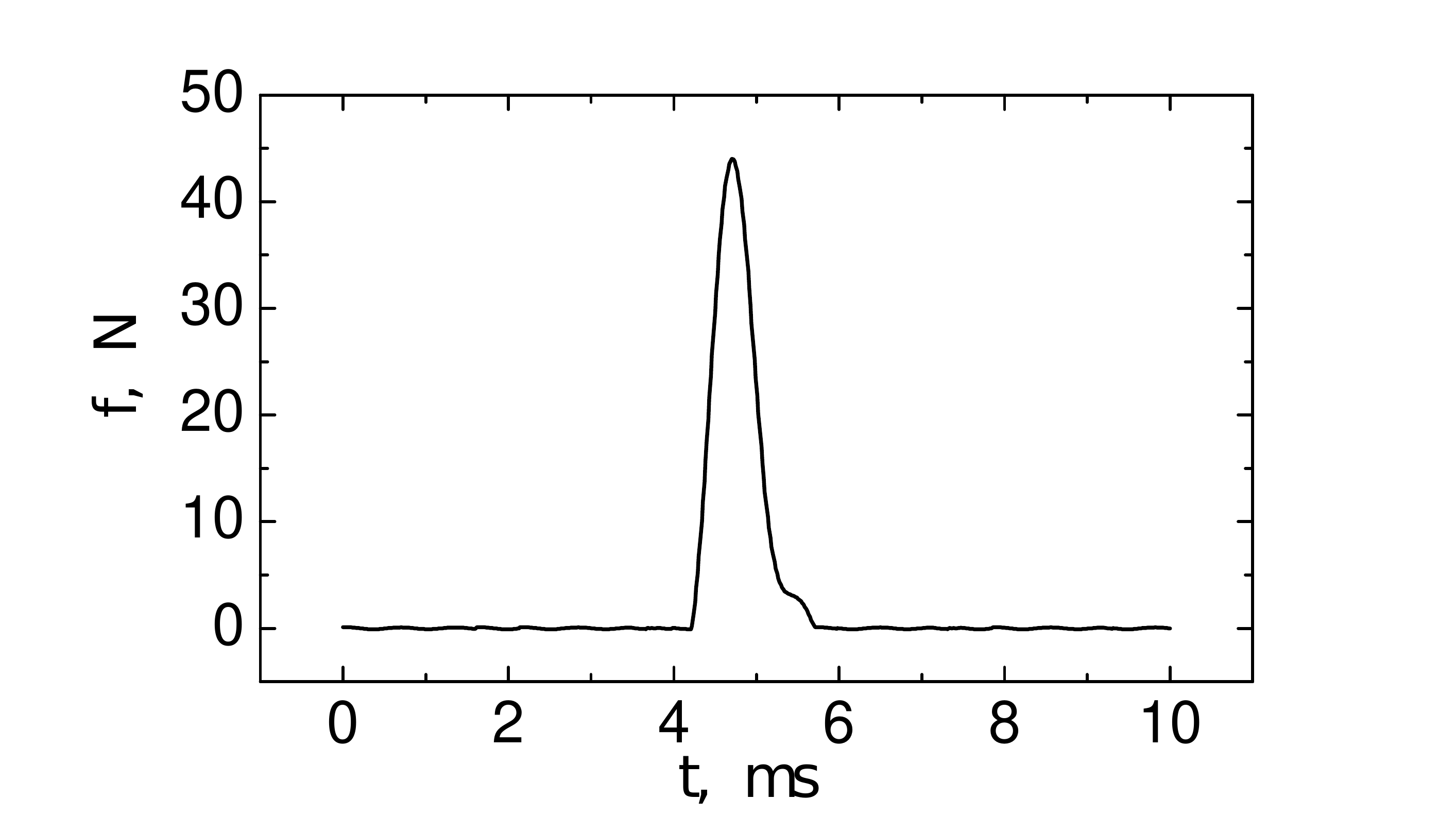}
\end{center}
\caption{The pulse shape of   signal generated by the impulse generator.} \label{mps:fig8}
\end{figure}
\begin{figure}
\begin{center}
\includegraphics[width= 7 cm, height= 7 cm]{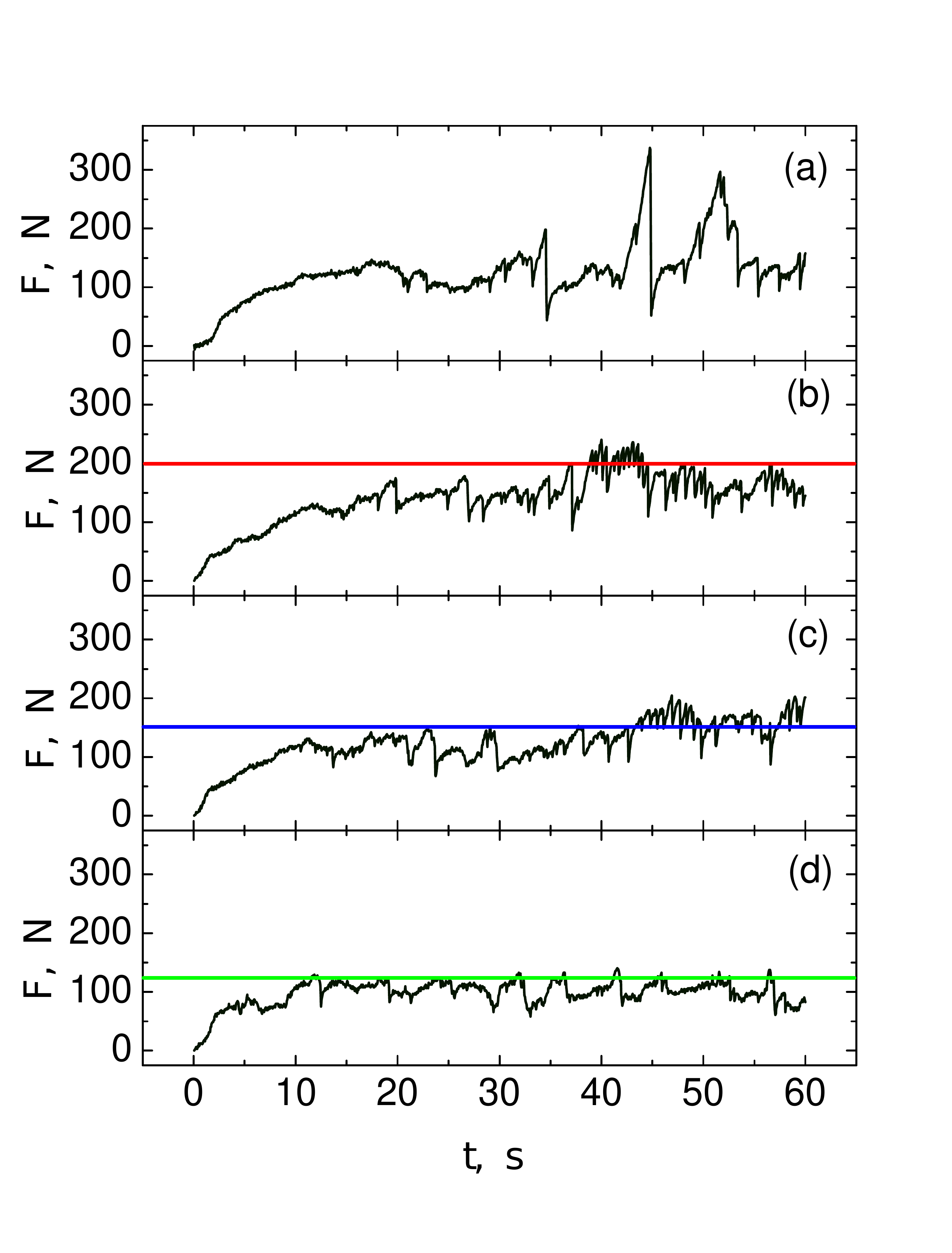}
\end{center}
\caption{Time dependencies of the traction force $F(t)$ during shear deformation of massif loaded by the force $P=60$N and stimulated by periodic signals of magnitude $f^m=44$N when it is chosen the different threshold  values: $F^\star=200$ N (b),  $F^\star=150$ N (c),  $F^\star=125$ N (d). For comparison,  the profile of $F(t)$ without external action (a).} \label{mps:fig9}
\end{figure}

 This feature of the parameter $\beta$ behaviour encourages  us to evaluate the Tsallis indexes for corresponding force  distributions. The application of the techique described above results in the graph plotted in Fig.~\ref{mps:fig7}b (bottom panel). We see that the mininum value of $q$ is reached at the same frequency $\Omega=500$~Hz.

The next series of experiments is concerned with studies of influences of external \emph{impulse} perturbations on the shear process. In this case the impulse generator (10) (the module (b), Fig.\ref{mps:fig1}) is used in order to  send the signal in the medium  when the traction force reaches some threshold value $F^\star$. The signals are sent with the frequency 1 Hz as long as the condition $F\geq F^\star$ is fulfilled.  Figure \ref{mps:fig8} shows the shape of one of the force impulses  generated by the generator. The amplitude of the pulse is $f^m=44$~N, lasted for $\tau =1$~ms. The granular medium  consists of 3000 cubes of the 10 mm sizes. In all experiments the medium is loaded with the weight of $P=60$~N.
The experiments are carried out at $F^\star=200,150,125$~N. 

The temporal dependencies of the traction force for these three threshold values are plotted in Fig.\ref{mps:fig9}, which also shows the  temporal  dependence without external action. From this figure  it follows that acting on the medium with external perturbations can avoid large tensions. 
It should be noted that the reduction of the traction force to a value smaller than the threshold value is not  achieved  by a single strike. This is well illustrated by Fig.~\ref{mps:fig9}b, c, where the perturbations are repeatedly sent in the medium. The experiments indicate that the use of such a mechanism makes the process of shear deformation smoother and that occurs at lower stresses.

These considerations are also acompanied by the Tsallis index evaluation. For the  threshold  values  $F^\star=\{0;125;150;200\}$N, it is obtained the indexes $q=\{0.591;  0.604; 0.603; 0.587\}$, which are  almost similar to each other.

\section{Concluding remarks}\label{mps:sec5}

Summarizing, it should be noted that experiments on the shear deformation of the granular medium exhibited  statistical similarity with natural seismic processes. The acoustic perturbations released by the granular medium during  shear deformation obey the power-like distribution similar to Gutenberg-Richter's law. It turned out that the power index lays in the range which is typical for natural  earthquakes. Moreover, for large acoustic disturbances the foreshocks and aftershocks were observed as well. It was shown that model aftershocks attenuate  in accordance with  the power law with the index close to 1, i.e.  Omori's law, that coincides with the statistical properties of  seismic process.

It is also shown that the sequence of energies of  acoustic disturbances is characterized by  the Tsallis index  $q = 0.516<1$.  This testifies about the presence of  long-range correlations in the shear motion and the system dynamics is defined by the mutual influence of a large number of rare events.

Our experiments with  different loads revealed the similarity of  processes of shear deformation, as well as for mixtures of grains of two sizes at different proportions of these grains number. The spectra of  traction forces are described by the power relations.

It is found out that the action by weak periodic waves with distinct frequencies on the granular medium which is in the state of shear deformation affect the medium's state. There exists the frequency ($\sim $500~Hz) when maximal effect of periodic stimulus is manifested.  The experiments have also shown that acting on the medium with small perturbations when the force of tension of a certain threshold value can be achieved, it is possible to attain smoother deformation. They also revealed that the smaller this threshold, the smoother the deformation.
 This indicates that the medium becomes smoother.

Finally, it should be noted that the detection of statistical regularities described by power laws testify that the granular medium evolves in the state of self-organized criticality. This state has quantitative characteristics which are similar to phenomena in seismic zones. 
This similarity can open new perspectives to control the seismically active zones due to the possibilities to affect the nonequilibrium block medium via the weak loading.

\end{document}